\documentclass[twocolumn,amsmath,amssymb,showpacs,superscriptaddress]{revtex4}
\usepackage{amssymb}
\usepackage{mathrsfs}
\usepackage{amsmath}
\usepackage{amsfonts}
\usepackage{graphicx}
\usepackage{subfigure}

\usepackage{graphicx}
\usepackage{dcolumn}
\usepackage{bm}

\begin{document}

\title{Enhanced Optical Conductivity Induced by Surface States in ABC-stacked Few-Layer Graphene}

\author{Jia-An Yan}
\affiliation{Department of Physics, Georgia Southern University, Statesboro, Georgia 30460 USA}
\author{W. Y. Ruan}
\affiliation{School of Physics, Georgia Institute of Technology, Atlanta, GA 30332 USA}
\author{M. Y. Chou}
\affiliation{School of Physics, Georgia Institute of Technology, Atlanta, GA 30332 USA}
\affiliation{Institute of Atomic and Molecular Sciences, Academia Sinica, Taipei 10617, Taiwan.}
\date{\today}
\begin{abstract}
The surface states of ABC-stacked few-layer graphene (FLG) are studied based on density-functional theory. These states form flat bands near the Fermi level, with the k-space range increasing with the layer number. Based on a tight-binding model, the characteristics of these surface states and their evolution with respect to the number of layers are examined. The infrared optical conductivity is then calculated within the single-particle excitation picture. We show that the surface states introduce unique peaks at around 0.3 eV in the optical conductivity spectra of ABC-stacked FLG when the polarization is parallel to the sheets, in good agreement with recent experimental measurement. Furthermore, as the layer number increases, the absorption amplitude is greatly enhanced and the peak position red-shifts, which provides a feasible way to identify the number of layers for ABC-stacked FLG using optical conductivity measurements.
\end{abstract}

\maketitle
\section{Introduction}

Monolayer graphene is a two dimensional (2D) system with linear dispersions near the K and K' points of the Brillouin zone (BZ) \cite{Novoselov2004,Berger2004,Novoselov2005,Zhang2005}. Low-energy charge carriers therein obey the Dirac-Weyl equation and behave like massless fermions.\cite{Geim2007,Neto2009} In few-layer graphene (FLG), the interlayer coupling introduces perturbations to the low-energy band dispersions. Consequently, the linear $\pi$ and $\pi^*$ bands near the Fermi level in monolayer graphene are modified in FLG, showing strong dependence on the stacking sequence as well as the layer number.\cite{Neto2009,Latil2006,Guinea2006, Lu2006, Aoki2007,Wang2007} These characteristic features make FLG attractive for practical nanoelectronics and optoelectronics applications. Indeed, recent experiments have shown that gated bilayer graphene exhibits a tunable band gap up to a few tenths of an eV,\cite{Ohta2006,Chen2008,Oostinga2008,Zhang2009,Mak2009} which is essential for practical device applications. Furthermore, intriguing physical properties, such as quantum Hall effect, Berry's phase,\cite{McCann2006,Koshino2009} chirality symmetry,\cite{Min2008} and optical conductivity \cite{Min2009} in FLG have been investigated.

The majority of natural graphite takes the Bernal stacking sequence (AB stacking), while only a small portion of natural graphite takes the rhombohedral ABC form.\cite{Lipson1942} AB-stacked FLG is believed to be thermodynamically stable. However, recent experiments show that ABC-stacked FLG can be obtained by mechanical exfoliation \cite{Mak2010, Lui2010} and by epitaxial growth on the SiC substrate.\cite{Wataru2010} These findings make the ABC-stacked FLG readily accessible in experiment.

As compared to the usual Bernal stacking, ABC-stacked FLG exhibits unusual band dispersions near the Fermi level. Specifically, the low-energy bands near the Fermi level in ABC-stacked FLG are surface states with their wavefunctions distributed on either $\alpha$- or $\beta$ atoms of the outermost layers. Consequently, no inter-atomic hopping is allowed in these states and the low-energy quasi-particle has an infinite mass. \cite{Guinea2006, Manes2007} In addition, ABC-stacked FLG is fundamentally interesting with the chirality of the charge carriers completely different from that in the monolayer and AB-stacked FLG.\cite{Zhang2010} Despite of previous theoretical studies on the electronic structure of FLG, the evolution of the surface bands of ABC-stacked FLG has not been well elucidated. Moreover, the surface states may have important effects on the optical transitions. In fact, the optical conductivity for the ABCA tetralayer is found to be distinct from the AB-stacked tetralayer.\cite{Mak2010} Therefore, understanding the origin of the surface states and their effects would be of great interest for further exploring ABC-stacked FLG.

In this work, we performed first-principles calculations to investigate the evolution of the surface states and their effects on the optical conductivity of ABC-stacked FLG. In Section II, we present the results of the band structure for ABC-stacked FLG. In Section III, a tight-binding model is developed to show the evolution of the surface bands with respect to the layer number. We discuss the effect of the surface bands on the infrared optical conductivity in Section IV. We find that the surface states introduce unique peaks in the infrared optical conductivity. Specifically, the peak position at around 0.3 eV is strongly dependent on the layer number and shifts to lower energies as the layer number increases. Interestingly, the optical transitions are only allowed for the polarization parallel to the graphene plane.

\section{Surface States in ABC-stacked Few-Layer Graphene}


\begin{figure}[tbp]
  \includegraphics[width=8.0cm]{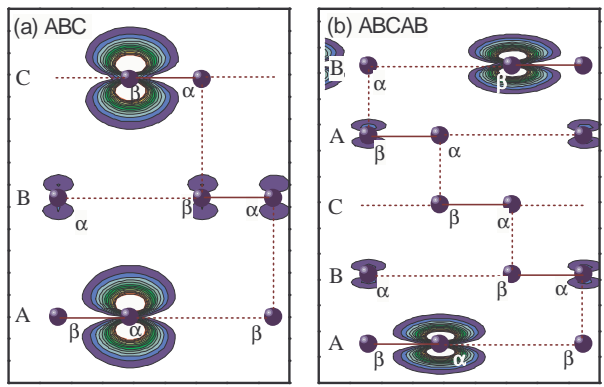}
 \caption{(Color online) Charge density distribution for the surface states in (a) the ABC trilayer and (b) ABCAB five-layer graphene. The charge densities associated with the states in the valence band at the $k$-points as indicated in Fig.~2 are plotted in the (110) plane.}\label{fig:cd}
\end{figure}


\begin{figure*}[tbp]
   \includegraphics[width=12cm]{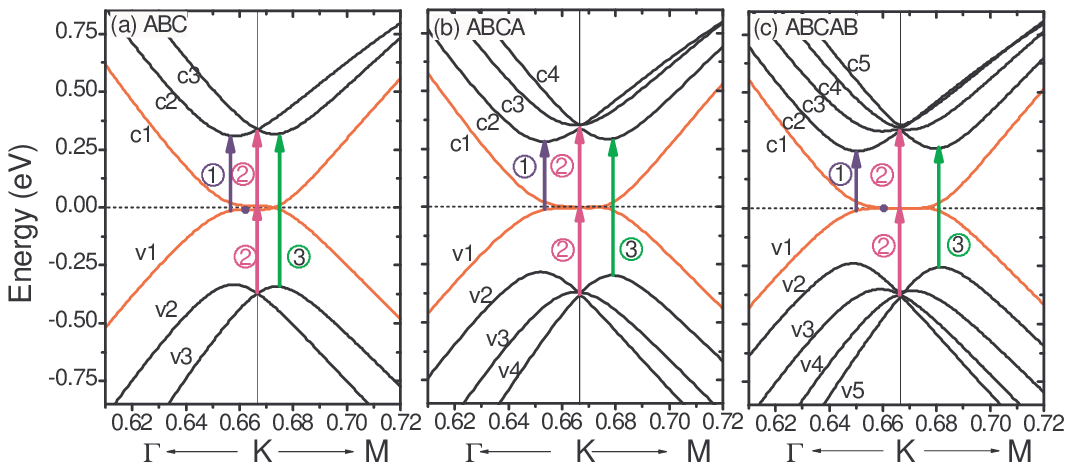}
 \caption{(Color online) Band dispersions for (a) ABC, (b) ABCA, and (c) ABCAB few-layer graphene in the vicinity of the Fermi level. The surface states are the flat bands near the Fermi level. The valence and conduction bands are labeled. The main allowed optical transitions between different sets of valence and conduction bands are indicated by arrow. For clarity, transition 1 between v2 and c1 is not marked.}\label{fig:bs}
\end{figure*}

In monolayer graphene, carbon atoms form a honeycomb structure with two inequivalent atoms ($\alpha$ and $\beta$) per hexagonal unit cell. An $\alpha$ atom has three nearest-neighbor $\beta$ atoms with relative displacements $\vec{\tau}_i$ ($i$=1,2,3). In ABC-stacked FLG, three different types of layers: $A$, $B$ and $C$ are present. Displacing an $A$-layer by $\vec{\tau}_1$ yields a $B$-layer, while further displacing a $B$-layer by $\vec{\tau}_1$ gives a $C$-layer. Thus the carbon atoms from different layers form a ladder structure. This feature can be identified clearly in the $\{110\}$ plane as shown in Fig.~\ref{fig:cd}. The unpaired carbon atoms at the two ends of the ladder induce interesting surface states, as will be discussed below.

First-principles calculations are performed to obtain the energy band structure for ABC-stacked FLG with layer number $L$ = 3--6 using the VASP code \cite{Kresse1993} with the projector-augmented wave (PAW) method and the local density approximation (LDA). The energy cutoff is set to be 500 eV. The Monkhorst-Pack $k$-point sampling of 36$\times$36$\times$1 and a large supercell with a $11$ \AA~vacuum region in the stacking-direction ($z$-direction) are used. Atoms in a supercell are fully relaxed until the force on each atom is less than 0.01 eV/\AA. The optimized interlayer separation is $d=3.32$ \AA~for ABC stacking configurations, which agrees well with results from previous first-principles calculations.\cite{Latil2006,Aoki2007}

Figures~\ref{fig:bs}(a)-(c) show the low energy dispersions for the ABC trilayer, ABCA tetralayer, and ABCAB five-layer graphene, respectively. The low-energy electronic states of $L$-layer FLG consist of $L$ pairs of conduction and valence bands. For ABC-stacked FLG, these bands split into two groups. First, one pair of bands ($v1$,$c1$) are located near the Fermi level. A prominent feature is that these two bands are quite flat in the vicinity of $K$. The flat region becomes broader as the number of layers increases. As shown in Fig.~\ref{fig:bs}, the other $L$--1 pairs are split-off bands which shift away from the Fermi level and pass through two points of around $\pm0.36$ eV at $K$. As will be discussed below, this energy reflects the interlayer hopping parameter $t_2$ in the tight-binding model. The two bands ($v2$, $c2$) are the so-called ``wizard-hat" bands. Due to the lower crystalline symmetry, the extrema of the wizard-hat bands move away from the $K$ point as indicated in Fig.~\ref{fig:bs}.


\begin{figure}[tbp]
  \includegraphics[width=9cm,clip]{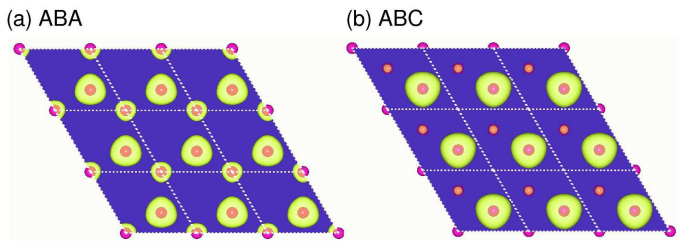}
 \caption{(Color online) Simulated constant-current STM images for (a) ABA and (b) ABC trilayer graphene. Red balls denote carbon atoms and yellow surfaces indicate isosurface of the charge density 1.0$\times$10$^{-4}$ $e/\AA^3$. The unit cells are indicated by dashed lines. The bias voltage is 0.25 V. }\label{fig:stm}
\end{figure}

Figures~\ref{fig:cd} presents the charge density contour for the flat bands ($v1$) at the k-point as indicated by a black dot in Fig.~\ref{fig:bs}. Indeed, the charge densities are mainly localized on the two outermost layers, and decrease exponentially into inner layers. These bands describe surface states in ABC-stacked FLG. Furthermore, the charge density distribution on the outermost layers concentrates on only one single sublattice. This feature will have important impact on the allowed optical transitions between the surface states and other bands.

Due to the different band dispersions between AB-stacked and ABC-stacked FLG, it would be possible to directly detect the surface states in ABC-stacked FLG using scanning tunneling microscopy (STM). In Fig.~\ref{fig:stm}, we present our simulated STM images for ABA and ABC trilayer graphene with a bias voltage of 0.25 V. In AB-stacked FLG, both sublattices ($\alpha$ and $\beta$) exhibit finite charge distributions, while for ABC-stacked FLG, the charge densities are localized on one sublattice in the surface layer. This is due to the fact that in AB-stacked FLG with an odd number of layers,\cite{Wang2007} the linear bands similar to that in monolayer graphene will be present. As a result, the STM image shows the charge distributions on both sublattices. In contrast, localized surface states in ABC-stacked FLG give rise to a distinct $\sqrt{3}\times\sqrt{3}$ STM pattern in the low voltage range of $\pm$0.3 V.

\section{Tight-binding Model for ABC-stacked Few-Layer Graphene}

The features of energy bands of ABC-stacked FLG can be clearly explained using the tight-binding model with the nearest-neighbor (NN) intralayer and interlayer interactions. The corresponding Hamiltonian (spin-unpolarized) is:
\begin{equation}\label{eq1}
\mathscr{H}=t_1\sum_{j\nu}\sum_{j'\nu'}
{}^{\prime}\hat{c}^{\dag}_{j\nu}\hat{c}_{j'\nu'}
     + t_2\sum_{j\nu}\sum_{j'\nu'}{ }^{\prime\prime}\hat{c}^{\dag}_{j\nu}\hat{c}_{j'\nu'},
\end{equation}
where $\sum'\left(\sum''\right)$ is a sum over all the intralayer (interlayer) NNs. $\hat{c}^{\dag}_{j\nu}$ is the creation operator for the $2p_z$ state localized on the $\nu$-th atom in the $j$-th
unit cell with a position vector ${\bf R}_j + \vec{ \xi}_\nu$.
The hopping parameters $ t_1=-3.0$ eV and  $t_2=0.36$ eV give
a reasonable fit to our DFT band data for trilayer graphene. Applying the
Fourier transform
\begin{eqnarray}\label{eq2}
\hat{c}_{j\nu} = \int_{\Omega_B} \frac{d^2k}{(2\pi)^2} e^{i{\bf
k}\cdot({\bf R}_j+\vec{\xi}_\nu)}\hat{c}_\nu({\bf k})
\end{eqnarray}
in Eq.(\ref{eq1}), the eigenvalue equation  $\mathscr{H}\Psi=E\Psi$ is equivalent to
the recurrence relations
\begin{equation}\label{eq3}
t_2{\bf h}^{(2)}_{l,(l-1)}C_{l-1} +t_1{\bf h}^{(1)}C_l +t_2{\bf
h}^{(2)}_{l,(l+1)}C_{l+1}=E C_l,
\end{equation}
where $C_l =(u_{\alpha,l}, u_{\beta,l})^T$ with $l = 1, 2, 3, ..., L$, and
$u_{\alpha,l}$($u_{\beta,l}$) are wave amplitudes on atom $\alpha$
($\beta$) in the $l$-th layer.  ${\bf h}^{(1)}$ is  given by

\begin{equation}\label{eq4}
{\bf h}^{(1)}=\left(
    \begin{array}{cc}
    0            &  f({\bf k})\\
     f({\bf k})^{*} & 0 \\
    \end{array}
   \right),
\end{equation}
where $f({\bf k})\equiv
\sum_{i=1}^{3}{e^{i{\bf{k}}\cdot{\bf{\tau}}_i}}\approx
-\frac{\sqrt{3}}{2}aq e^{-i\varphi_q}$ for small $q$ (${\bf q}\equiv
{\bf k} -{\bf K})$.

For ABC-stacked FLG, we have

\begin{equation}\label{eq5}
{\bf h}^{(2)}_{l,(l-1)}=\left(
    \begin{array}{cc}
     0   &  0\\
     1   &  0 \\
    \end{array}
   \right),
{\bf h}^{(2)}_{l,(l+1)}=\left(
    \begin{array}{cc}
     0   &  1\\
     0   &  0 \\
    \end{array}
   \right).
\end{equation}
For ABC-stacked graphite, Eq.~(\ref{eq3}) yields Bloch-type solutions with $C_l\sim e^{iqld}$. These can be used to construct the standing wave solutions for FLG with a finite layer number $L$. Detailed solutions to Eq.~(\ref{eq3}) for an arbitrary $L$ value will be published elsewhere. Here we focus on the surface states in ABC-stacked FLG.

Essentially, the geometrical ladder structure of an ABC-stacked FLG can be mapped to a linear atomic chain with two atoms per unit cell. For a finite $L$-layer ABC stacking, the general solutions to Eq.~(\ref{eq3})
are superposition of the two traveling waves with wave vectors $\pm q$. Possible values of $q$ (and $E$) are determined by the boundary conditions
which can be expressed as:\cite{Tsuji1960}
\begin{equation}\label{eq6}
\texttt{sin}[(L+1)qd]+\frac{t_2}{t_1|f(\bf{k})|} \texttt{sin}(Lqd)=0,
\end{equation}
with $0\le qd\le\pi$. When $|t_2/[t_1 f({\bf k})]|<1+1/L$, Eq.~(\ref{eq6})
has $L$ real roots $q_i$ ($i$=1, 2, ..., $L$) satisfying $q_id \in
[0,\pi]$.



When $|t_2/[t_1 f({\bf k})]|>(1+1/L)$, only $L$--1 real roots are
obtained from Eq.~(\ref{eq6}) for $qd \in [0,\pi]$. The lost root can be
recovered by assuming a complex wave vector with $qd=\pi+i\theta$
($\theta>$0) and $\theta$ satisfying
\begin{equation} \label{eq7}
\texttt{sinh}[(L+1)\theta]+t_2/(t_1 |f({\bf k})|)
\texttt{sinh}(L\theta) =0.
\end{equation}
The corresponding energy is
\begin{equation}\label{eq8}
E = \pm \sqrt{t_1^2 |f({\bf k})|^2+t_2^2 +2t_1 t_2 |f({\bf k})|
\texttt{cosh}{\theta}}.
\end{equation}
Since the k-points involved are close to K, the energy values in Eq. (8) are quite small and weakly dependent on k. This gives rise to flat band dispersions near the K point.

\begin{figure}[tbp]
  \includegraphics[width=7.0cm]{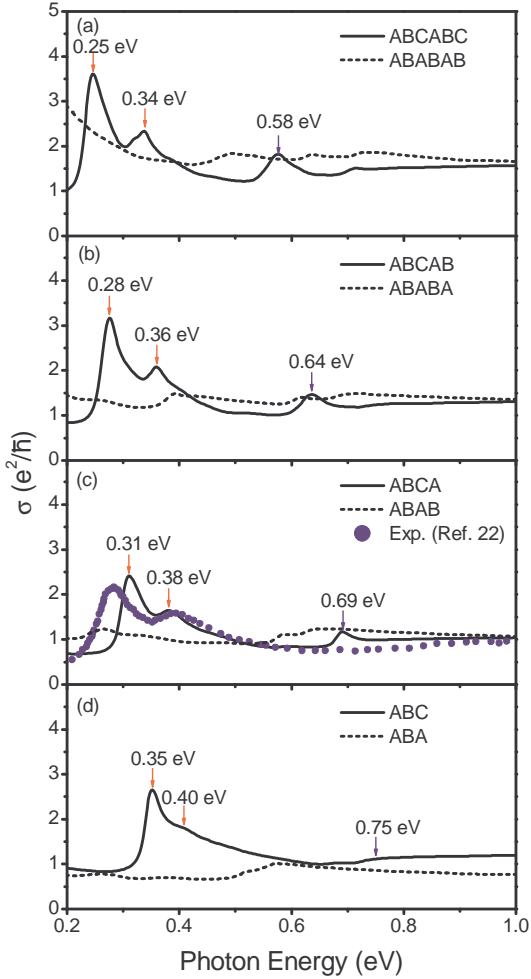}
 \caption{(Color online) Calculated optical conductivity for ABC-stacked few-layer graphene with the light polarization parallel to the graphene sheets. Results of the AB-stacked graphene (dashed lines) are also shown for comparison. The peak positions are indicated by arrows. A 10 meV Gaussian broadening is used. }\label{fig:cond1}
\end{figure}

\section{Optical Conductivity of ABC-stacked Few-Layer Graphene}
The optical properties, especially the infrared absorption spectra of AB-stacked FLG have been thoroughly studied by many experimental and theoretical investigations.\cite{Lu2006-1,Casiraghi2007,Jiang2007,Yan2007,Li2008,Stauber2008,Li2009,Wang2009,Mak2009} In contrast, existing work on ABC-stacked FLG is relatively limited.\cite{Lu2006-2,Koshino2010} Below we report the calculated dynamical dielectric functions and optical conductivity of ABC-stacked FLG with $L$ = 3--6 from first-principles. The electronic structure of ABC-stacked FLG discussed above has important implications on the optical response. Here we will focus on the role of the surface states in the optical response of ABC-stacked FLG. Within the independent particle picture, the imaginary part of the frequency-dependent dielectric function can be calculated via:\cite{Gajdos2006}

\begin{eqnarray}
\varepsilon_{\mu \nu}^{(2)} (\omega)& = &\frac{4\pi^2 e^2}{\Omega} \lim_{q\to0} \frac{1}{q^2} \sum_{cvk} 2w_\textbf{k} \delta(\epsilon_{c\textbf{k}}-\epsilon_{v\textbf{k}}-\omega) \\ \nonumber
& & \times \langle u_{c\textbf{k}+\textbf{e}_\mu q}|u_{v\,\textbf{k}}\rangle\langle u_{c\,\textbf{k}+\textbf{e}_\nu q}|u_{v\textbf{k}}\rangle^*.
\end{eqnarray}
Here, $w_k$ is the weight of the k-point in the BZ summation, $u_c$ ($u_v$) and $\epsilon_{c\textbf{k}}$ ($\epsilon_{v\textbf{k}}$) are the wave functions and energies of the conduction (valence) bands, respectively. The factor of 2 accounts for the spin degeneracy. Based on the obtained $\varepsilon$, the optical conductivity can be calculated by using $\sigma=\omega \varepsilon^{(2)}/4\pi$, which can be compared directly with experiment.

Graphene and multilayer graphene have an almost zero energy gap. In order to calculate the dielectric functions accurately, we have used a very dense $k$-grid sampling of 210$\times$210 and checked the convergence. This k-grid yields well-converged peak positions and relative amplitudes of the dielectric functions.

\begin{figure}[tbp]
   \includegraphics[width=7.0cm]{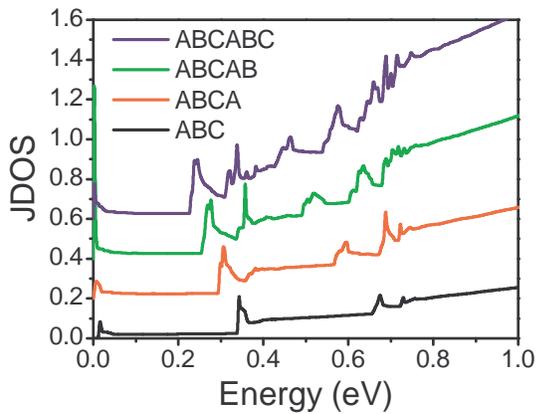}
 \caption{(Color online) Calculated joint density of states (JDOS) for ABC-stacked few-layer graphene. A 10 meV Gaussian broadening is used. For clarity, the curves have been relatively offset by 0.2 along the vertical direction.}\label{fig:jdos}
\end{figure}

In Fig.~\ref{fig:cond1}, we show the calculated optical conductivity for ABC-stacked FLG with $L$ = 3--6, with the polarization parallel to the graphene sheet. The calculated results for the corresponding AB-stacked FLG are also shown for comparison. Note that within the visible light range, the optical conductivity almost stays constant. \cite{Min2009} The corresponding joint densities of states (JDOSs) are presented in Fig.~\ref{fig:jdos}. In addition to the step-like singularities which are also observed in AB-stacked FLG, the JDOSs in ABC-stacked FLG show a 1D-like divergence varying as $1/\sqrt{E}$. This feature arises from the surface bands around the $K$ point of the Brillouin zone where the ``wizard-hat" ($c2$,$v2$) bands have their extrema, as previously discussed by Guinea {\it et al}.\cite{Guinea2006}

The above features lead to a strong response in the optical conductivity spectrum. For example, the strongest infrared absorption feature was found at 0.35 eV and 0.31 eV for the ABC trilayer and ABCA tetralayer, respectively. It corresponds to transition 1 ($v1 \rightarrow c2$ and $v2 \rightarrow c1$, one of which is denoted by blue arrows in Fig.~\ref{fig:bs}) at around 0.3 eV, where the JDOS diverges. In this transition, the electronic states in the vicinity of the flat region of v1 and c1 will couple to the extrema of band c2 and v2, respectively. The second prominent absorption feature is found at 0.40 eV and 0.38 eV for the ABC trilayer and ABCA tetralayer, respectively. These two transitions involve the surface bands, and exhibit a significant enhancement effect as the layer number increases. The third absorption peak, which is invisible for the ABC trilayer, is located at around 0.69 eV for the ABCA tetralayer. This peak arises from transition 2 between the extrema of the ``wizard-hat" bands (see Fig.~\ref{fig:bs}). In contrast, this absorption amplitude is much weaker, only about 54\% of the first absorption peak in the ABCA tetralayer.


As can be seen from Fig.~\ref{fig:cond1}, the absorption peaks evolve as the number of layers increases for ABC-stacked FLG. In particular, the position of the first peak of the ABCA tetralayer red-shifts by 40 meV compared with that of the ABC trilayer. The second and third absorption peaks also red-shift as the layer number increases.

The calculated optical conductivity can be compared with experiment. In Fig.~\ref{fig:cond1}(c), the measured infrared optical conductivity for the ABCA tetralayer\cite{Mak2010} is plotted together with our calculated result. Overall, the calculated optical conductivity spectra are in good agreement with experimental data. However, there are slight deviations between calculations and experiment. The calculated peak at around 0.31 eV is 0.05 eV higher than the observed value of 0.26 eV, while the second absorption peak at 0.38 eV is 0.03 eV higher than the experimental data of 0.35 eV. Furthermore, the third (weaker) resonance feature expected for the ABCA tetralayer around 0.67 eV is difficult to identify in the experiment.
Possible reasons for these differences may be ascribed to many-body interactions, exciton effects,\cite{Yang2009} or doping.

\begin{figure}[tbp]
  \includegraphics[width=7.5cm]{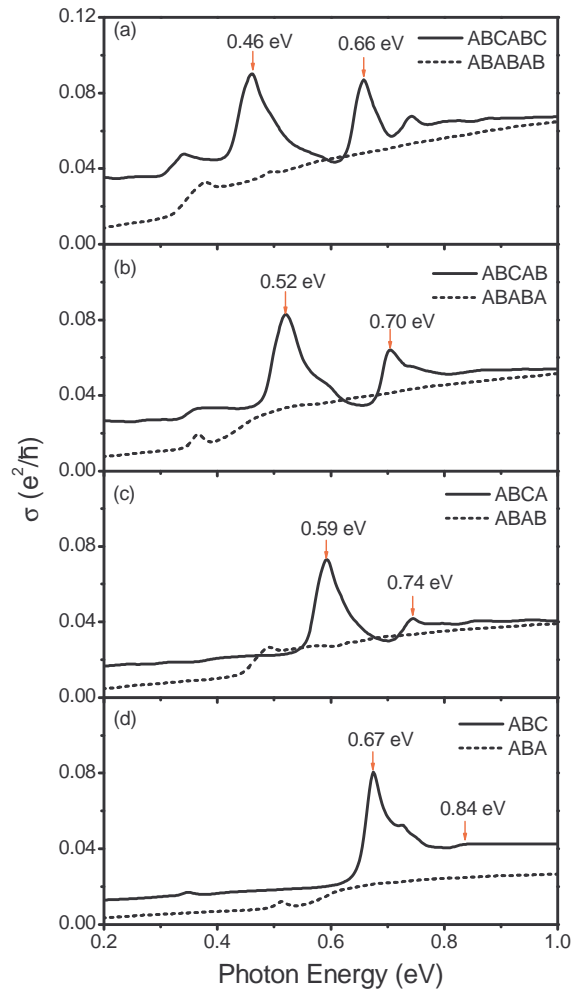}
 \caption{(Color online) Calculated optical conductivity for ABC-stacked graphene with the light polarization perpendicular to graphene sheets. Results of the AB-stacked graphene (dashed lines) are also shown for comparison. The peak positions are indicated by arrows. A 10 meV Gaussian broadening is used.}\label{fig:cond2}
\end{figure}

In Fig.~\ref{fig:cond2}, the optical conductivity for ABC-stacked FLG is presented with the polarization perpendicular to the graphene plane. The calculated results for the corresponding AB-stacked FLG are also shown as dashed lines. The local effect on the optical absorption spectra has not been considered in this work. The amplitude is more than one order of magnitude smaller than that shown in Fig.~\ref{fig:cond1}. This polarization effect is also observed for AB-stacked bilayer graphene,\cite{Yang2010} with or without electric field. Note that the peak positions are completely different as compared with those in Fig.~\ref{fig:cond1}. In particular, the allowed transitions for the surface bands disappear in Fig.~\ref{fig:cond2}. This result suggests that the optical transition between electronic bands are extremely sensitive to the polarization direction.


\section{Conclusions}

In summary, we have investigated the evolution of the surface states in ABC-stacked FLG and their effects on the optical conductivity. These surface states are localized on one single sublattice in the outermost layers, and the amplitudes decay exponentially into the inner layers. The surface bands are quite flat around $K$ in the Brillouin zone, and the flat region increases as the layer number increases. The formation of these surface states is well elucidated by a tight-binding model.

Compared with AB-stacked FLG, the surface states in ABC-stacked FLG have significant effects on the optical absorption spectra. These surface states introduce pronounced absorption peaks at around 0.3 eV. As the layer number increases, the absorption amplitudes are greatly enhanced due to the strong localization of the surface states, and the absorption peak red-shifts as the layer number increases. Interestingly, the absorption due to the surface states is sensitive to the polarization direction. The enhanced absorption can only be found for the polarization parallel to the graphene sheets.

\begin{acknowledgements}
J.A.Y. thanks Dr. Xiaojun Wang for useful discussions. We acknowledge the support by the US Department of Energy, Office of Basic Energy Sciences, Division of Materials Sciences and Engineering under Award No. DEFG02-97ER45632. Computational resources are provided by the National Energy Research Scientific Computing Center (NERSC).
\end{acknowledgements}

\end{document}